\documentclass[fleqn,usenatbib]{mnras}

\usepackage{mathrsfs}
\usepackage{newtxtext,newtxmath}

\usepackage[T1]{fontenc}
\usepackage{ae,aecompl}

\usepackage{graphicx}	
\usepackage{amsmath}	
\usepackage[colorinlistoftodos]{todonotes}

\definecolor{sjs}{rgb}{0.13, 0.54, 0.13}

\title[Impact of N(z) width error on cosmology]{The Impact of Tomographic Redshift Bin Width Errors on Cosmological Probes}

\author[I. S. Hasan et al.]{
Imran S. Hasan,$^{1}$ \thanks{E-mail: ishasan@ucdavis.edu}
Samuel J. Schmidt,$^{1}$
Michael D. Schneider$^{2}$
and J. Anthony Tyson$^{1}$
\\
$^{1}$Department of Physics, University of California, One Shields Ave., Davis, CA 95616, USA\\
$^{2}$Lawrence Livermore National Laboratory, Livermore, CA 94551, USA\\
}

\date{Accepted XXX. Received YYY; in original form ZZZ}

\pubyear{2021}

\begin{document}
\label{firstpage}
\pagerange{\pageref{firstpage}--\pageref{lastpage}}
\maketitle

\begin{abstract}
Systematic errors in the galaxy redshift distribution $n(z)$ can propagate to systematic errors in the derived cosmology. We characterize how the degenerate effects in tomographic bin widths and galaxy bias impart systematic errors on cosmology inference using observational data from the Deep Lens Survey. For this we use a combination of galaxy clustering and galaxy-galaxy lensing. We present two end-to-end analyses from the catalogue level to parameter estimation. We produce an initial cosmological inference using fiducial tomographic redshift bins derived from photometric redshifts, then compare this with a result where the redshift bins are empirically corrected using a set of spectroscopic redshifts. We find that the derived parameter $S_8 \equiv \sigma_8 (\Omega_m/.3)^{1/2}$ goes from $.841^{+0.062}_{-.061}$ to $.739^{+.054}_{-.050}$  upon correcting the n(z) errors in the second method.
\end{abstract}

\begin{keywords}
galaxies: distances and redshifts -- gravitational lensing: weak -- methods: observational
\end{keywords}



\section{Introduction}
The large scale structure (LSS) today is imprinted with the initial matter power spectrum, and by extension, the initial conditions of the universe. The statistical properties of the LSS and their time evolution are governed by the cosmic acceleration, matter-energy content, and gravity. Cosmology surveys leverage this sensitivity as a means to understand the fundamental nature of dark matter and dark energy. A prerequisite for many of these studies is knowledge of the redshift of observed galaxies. Surveys rely increasingly on photometric redshift estimates (photo-$z$) to satisfy this requirement. However, as data sets become photometrically deeper, and new probes are introduced, photo-$z$s have become a leading systematic error in cosmology inference.

Two point statistics, like galaxy clustering and galaxy-galaxy lensing, of photometrically observed galaxies have been recognized as powerful techniques for characterizing the LSS. Galaxies are divided into redshift bins, which enables measurements of time evolution of the LSS \citep{andy, zhan, growth}. This requires sufficiently accurate redshifts for every galaxy used. Obtaining a complete spectroscopic redshift sample to complement the photometric data is intractable. Photometric samples are dominated by many millions of faint galaxies, for which obtaining high signal to noise spectroscopic redshifts (spec-$z$s) is prohibitively costly. Instead, photo-$z$ algorithms are combined with much smaller samples of spec-$z$s to generate redshift estimates. The spectroscopic samples are often photometrically non-representative of the galaxies for which photo-$z$s are measured, and are re-weighted to correct for this \citep{hscpz}. 

These two point statistics provide an enhanced measure of structure if used independently. This enhancement is often modeled as a linear factor in the amplitude of density perturbations, $b$, the galaxy bias. If galaxy-galaxy lensing and galaxy clustering are used together in a joint analysis, however, one can solve for the linear galaxy bias and obtain an unbiased measure of structure formation.

Galaxy-galaxy lensing is the correlation of the shape of background galaxies with the position of foreground galaxies \citep{2013MNRAS.432.1544M}. The mass from the foreground objects act as a magnifying glass, distorting light from sources behind them into a coherent tangential alignment. Galaxy-galaxy lensing involves one factor of the galaxy bias, $b$, as the position of luminous foreground objects are targeted. 

Galaxy clustering is the correlation between the galaxy positions \citep{2013MNRAS.432.1544M}. Because of the galaxy-halo connection, and because galaxies reside in the same LSS, their positions have non-zero correlation. Galaxy clustering involves two factors of the galaxy bias, as it correlates the positions of two galaxies. Galaxy clustering can be combined with galaxy-galaxy lensing in order to constrain the galaxy bias, and produce cosmological constraints.

Galaxy clustering and galaxy-galaxy lensing primarily probe the matter field, and are consequently most sensitive to $\Omega_m$ and $\sigma_8$, the matter content of the universe and the root mean square of mass fluctuations respectively. Sensitivity to these parameters is degenerate; they are often combined to define $S_8 = \sigma_8(\Omega_m/.3)^{\alpha}$ to break the strong correlation between $\Omega_m$ and $\sigma_8$ in these experiments, where $\alpha$ is optimally selected to break the degeneracy.

The additional probes are accompanied by additional systematic errors. Typically, one uses a cosmological model that treats the galaxy bias and error in mean redshift of each tomographic bin as nuisance parameters and marginalizes over them \citep{abbot, 2018arXiv180901669T, 2020arXiv200715632H}. However, systematic sources of error that are degenerate with galaxy bias and mean redshift error which are omitted from the model introduce degenerate solutions. Specifically, errors in higher order moments of photo-$z$ distributions, e.~g.~width, can imitate error in the galaxy bias. In a scenario where errors in photo-$z$ widths are unaccounted for, the imprint they leave on measured two point statistics can erroneously be fit by an evolving galaxy bias. If left unaccounted for, such degeneracies have the potential to bias the cosmological parameter estimation. Non-linear galaxy bias is another potential confounding factor.

The need for control in systematics comes into sharper focus when we consider the results of weak lensing surveys alongside other cosmology experiments. A complete cosmological model must be able to accurately describe the universe across epochs. It is therefore interesting to compare constraints resulting from measurements of the Cosmic Microwave Background (CMB) at $z \sim$ 1100 to dark matter and dark energy probes from $z \sim$ 1.5 and lower. $S_8$ predictions from \emph{Planck}, the highest precision CMB experiment to date, are 2-3$\sigma$ above those of several weak lensing results \citep{2020A&A...641A...6P, 2020arXiv200715632H}.  

This $S_8$ tension may be providing a hint of new physics beyond the standard $\Lambda$CDM cosmological model \citep{2018LRR....21....2A}. However, interpretation of these 2-3 $\sigma$ tensions as a departure from the concordance model must be tempered by discretion. Weak lensing surveys suffer from a myriad of pernicious systematic effects, and the $S_8$ tension may be due entirely to these. Therefore, high precision measurements demand robust mitigation of systematic errors as a necessary prerequisite if they are to address the $S_8$ tension, or test cosmological models more broadly.   Systematic effects in experiments are often the result of tails of sample distributuions disagreeing with the assumed model.

Control and mitigation of systeamtics in weak lensing surveys-on many fronts-has matured in recent decades. For example, advancements in the design of observational facilities, observation techniques, galaxy shape measurement algorithms, point spread function (PSF) measurement and modeling, and CCD physics have enabled high precision cosmology results in completed Stage II and ongoing Stage III experiments. However, treatment of photo-$z$ errors and systematics have, in general, not matured at the same pace. Furthermore, photo-$z$ performance may have been acceptable for previous and current weak lensing results, but the improving quality of data create much more stringent requirements for photo-$z$ performance. This is especially true for experiments designed to test the time evolution of dark energy and alternate dark energy models more broadly, which require systematic errors to be controlled at the sub-per cent level \citep{2015APh....63...81N} 

In this work, we investigate the impact that realistic errors in the galaxy redshift distribution of a tomographic bin sample, the n(z), in a weak lensing experiment can have on cosmological inference from combining galaxy clustering and galaxy-galaxy lensing.
Stress testing n(z) errors on real data of similar quality to ongoing and future experiments can probe failure modes and error regimes that may not present themselves in simulations, or in an analytic treatment. 

We utilize a Rubin Observatory \citep{2019ApJ...873..111I} Legacy Survey of Space and Time (LSST) \citep{2002SPIE.4836...10T} precursor survey, the Deep Lens Survey \citep{dave} 
as a setting to evaluate the degree to which n(z) and galaxy bias estimation errors influence cosmology inference. While error effects due to placing galaxies in incorrect tomographic bin 
and error on n(z) mean have been considered in elsewhere \citep{2018arXiv180901669T}, here we explicitly examine the extent to which errors in n(z) in tomographics bins, which can be degenerate with galaxy bias errors, propagate to systematic errors in the inferred cosmological parameters.

This paper is structured as follows. In \S \ref{sec:twopointprobes} we formally define the two angular two-point statistics which will be used as cosmological probes. We will indicate explicitly the potential for n(z) errors to be conflated with galaxy bias in doing so. In \S \ref{sec:data} we introduce the Deep Lens Survey, a Rubin Observatory LSST precursor weak lensing survey. In \S \ref{sec:meas} we present measured two point statistics from our data set. In \S \ref{sec:correct_nz} we outline empirical corrections to the n(z) bins, using a validation spectroscopic redshift data set. 
We present the formalism for our cosmology inference using our measured two point statistics in \S \ref{sec:cosmo}. In \S \ref{sec:results} we lay out the results of our study, examining the ultimate effect our adjustments to the estimated n(z) shape has on cosmological inference, and in \S \ref{sec:compare_cosmo} we compare the results of our cosmology inference with other studies. We conclude the paper in \S \ref{sec:conclusion} and look to future work.

\section{Two Point Probes}
\label{sec:twopointprobes}
The probes used in this analysis form auto or cross correlations of tomographic bins of the projected angular galaxy field and the projected angular convergence field. We will adopt a notation where we use $\delta_g^i$ to refer to the $i$th tomographic bin of the galaxy field, and $\kappa^j$ where $\kappa$ refers to the $j$th tomographic bin of the convergence field.  We assume the galaxy field, $\delta_{g}$, is a biased representation of the matter field, $\delta_{m}$, and that we can parametrize this bias with a linear correction: $b\delta_{m}$ = $\delta_{g}$, where $b$ is the galaxy bias. The galaxy field is sampled by observing galaxies with redshift distributions, $p(z) = p(\chi)(d\chi /dz)$ where $\chi$ is the line of sight comoving distance.
We adopt the notation of using $p^i_{\delta_g}(\chi)$ and $p^i_{\kappa}(\chi)$ to represent the normalized redshift distribution of the $i$th tomographic bin of the galaxy and convergence field, respectively. Because the non-linear matter power spectrum, $P(k,z)$, with $k$ the Fourier space wave vector and redshift $z$, gives rise to these fields, measuring them constrains $P(k,z)$, and cosmological parameters by extension. 

Under the flat sky Limber approximation, the Fourier space wave vector $k \xrightarrow{} \frac{l + 1/2}{f(\chi)}$ and $P(k,z) \xrightarrow{} P\bigg[\frac{l + 1/2}{f(\chi)},z\bigg]$ where $f(\chi)$ is the comoving angular diameter distance.

The galaxy-galaxy lensing power spectrum is given by  

\begin{equation}
    C^{ij}_{g\kappa}(l) = b^{i}\int_{0}^{\chi_h} \frac{q^{i}_{\kappa}(\chi)p_{\delta_g}^{j}(\chi)}{a(\chi)f(\chi)} P\bigg[\frac{l + 1/2}{f(\chi)},\chi\bigg] \, \mathrm{d}\chi 
	\label{eq:ggl}
\end{equation}

Here, $\chi_h$ is the line of sight comoving horizon distance, $a$ is the scale factor, and $q^{i}$ is the lensing efficiency of the $i$th tomographic bin, given by

\begin{equation}
    q^{i}(\chi) = \frac{3H_{0}^{2}\Omega_{m}}{2c^2} \int_{\chi}^{\chi_h} p^{i}_{\kappa}(\chi')\frac{f(\chi' - \chi)}{f(\chi')} \, \mathrm{d}\chi' 
	\label{eq:eff}
\end{equation}
Where $p^{i}_{\kappa}(\chi)$ is the normalized redshift distribution of galaxies in the $i$th tomographic bin of the convergence field, $c$ is the speed of light, and $H_0$ is the Hubble Constant. Additionally, here $p^{i}_{g}(\chi)$ is the normalized redshift distribution of galaxies in the $i$th tomographic bin of the galaxy field, and $b^i$ is the linear galaxy bias in the $i$th bin.

The galaxy clustering power spectrum is given by

\begin{equation}
    C^{ij}_{gg}(l) = b^{i}b^{j}\int_{0}^{\chi_h} \frac{p^{i}_{\delta_g}(\chi)p^{j}_{\delta_g}(\chi)}{f^2(\chi)} P\bigg[\frac{l + 1/2}{f(\chi)},\chi\bigg] \, \mathrm{d}\chi 
	\label{eq:cluster}
\end{equation}

the linear bias $b^i$ in the galaxy-galaxy lensing power spectra is assumed to be identical to that in the galaxy clustering spectra (e.g. $r^{i} = b^{i}_{\times}/b^{i} = 1$).

We emphasize that in equations \ref{eq:cluster} and \ref{eq:ggl}, the power spectra amplitude depend on both the galaxy bias \emph{and} the $n(z)$ distribution of the galaxies being considered. The covariance in these quantities may have a detrimental effect on
the accuracy of any fit to a cosmology model if not captured correctly. $n(z)$ models are required as input for cosmology inference. Although it is true the means of their distributions are treated as a nuisance parameter and marginalized over, this approach may not capture the impact that higher order moments in the $n(z)$ impart on the resulting estimates.

To help motivate this argument, we consider a toy example. Suppose the true $n(z)$ of a tomographic bin is broader than the estimated $n(z)$. We proceed, calculating a model $w(\theta)$ auto-correlation using the estimated $n(z)$ as input. The model signal amplitude will rise as galaxy positions in narrower redshift slices have stronger correlations, as galaxies in narrower redshift slices are more likely to reside in the same LSS and have correlated positions on the sky.
This will have a down-stream effect in model fitting; the fitted galaxy bias will produce a smaller value than the true galaxy bias, as the predicted model is already overestimated.

The degenerate effect the galaxy bias and errors in the shape of the $n(z)$ create multiple, degenerate solutions to cosmology inference, potentially creating important but unaddressed sources of error. As a reminder, this is our primary concern in this work. In section \ref{sec:cosmo} we will ultimately consider the degree to which unmitigated $n(z)$ errors ultimately impact cosmology inference. 

In practice, the measured data will be in real space, which will require us to transform the power spectra. The angular two point correlation function, $w(\theta)$, and tangential shear, $\gamma_t$ respectively, are given by 

\begin{equation}
    w(\theta) = \sum_l \frac{2l +1}{4\pi}P_l(\mathrm{cos}(\theta))C_{\delta_g \delta_g}(l)
\end{equation}
\begin{equation}
    \gamma_t(\theta) = \frac{1}{2\pi} \int_0^\infty J_{2}(l\theta) C_{\delta_g\kappa}(l) l \, \mathrm{d}l
\end{equation}

Where $P_l(x)$ is the Legendre polynomial of order $l$ and $J_{n}(l\theta)$ is the $n$th order Bessel function of the first kind. For this analysis the real space correlation functions are measured with estimators. The measurement uses observed galaxies as tracers of the galaxy field and convergence field. A full discussion of the estimators follows in section \ref{sec:meas}. 

\section{Data}
\label{sec:data}
The Deep Lens Survey \citep[DLS,][]{dave} is a weak lensing survey designed as a precursor to the Rubin Observatory LSST survey \citep{2002SPIE.4836...10T}. Broadband imaging was obtained in four broad-band filters-$BVRz'$-to enable photo-$z$ estimation and shape measurement. As a result, it provides a staging ground to examine the impact photo-$z$ estimation has on determination of galaxy bias and cosmology constraints.

The DLS consists of five 2 square degree fields, distributed across the sky with wide angular separation between them. As a weak lensing survey, the fields were chosen to avoid the Milky Way plane and bright low-z galaxies, and are otherwise chosen without regard to known structures. The total area of the footprint was chosen such that deep photometric data could be achieved for shape measurement and photometric redshift estimation. The area was evenly split into five widely separated fields to avoid cosmic variance. Each field (F1-F5) was subdivided into a 3x3 grid of 9 sub-fields which were observed with a shift and stare dithering pattern. Observations for the R band were only carried out in the best seeing conditions (with a typical full width half max of .9 arcseconds), accumulating 18000s in exposure time at any point in the survey. The $R$ band exposure time enables a typical 5$\sigma$ detection limit of $m_R = 27$ for point sources. Observations in the remaining filters were 12000s each, and the seeing conditions in which they were observed was prioritized in the following order: $V$, $B$, $z'$. Because priority was given to the $R$ band in exposure time and seeing, detection and shape measurement are carried out on the R band coadded images \citep{dave}.

F1 and F2 lie in the northern hemisphere, while the remaining fields, F3, F4, and F5 lie in the southern hemisphere. Observing was carried out with the MOSAIC I camera at Kitt Peak National Observatory (KPNO) on the Mayall 4-m Telescope and at Cerror Tololo Inter-American Observatory (CTIO) with the MOSAIC II camera on the Blanco 4-m Telescope.
The focal planes in both cameras have a 4x2 array of 2k x 4k CCDs with .25 arcsecond pixel scale, spanning a 35 square arcminute field of view. 

The initial photometry estimation was carried out using \texttt{Source Extractor} MAG\_AUTO outputs \citep{ba}. Photometric calibration was performed on magnitudes and colours using \texttt{COLORPRO} \citep{2006AJ....132..926C}
to correct for variations in seeing conditions between visits. A global linear least-squares (\texttt{\"Ubercal}) approach in the style of the Sloan Digital Sky Survey (SDSS) \citep{nik} was implemented, and subsequently mitigated spatial variations in photometry \citep{daveuber}. Galactic extinction was corrected using the Schlegel reddening maps \citep{sch}.

\subsection{Shape measurement}
\label{sec:shape} 
Seeing conditions from the atmosphere, dome, telescope optics, and detector which define the point spread function (PSF) impart ellipticities to measured sources. To recover the true ellipticities of galaxies, these effects must be removed. Removal of these signatures in the shape catalogue used in this analysis are fully described in \cite{dls2d} and we briefly summarize them below. 

The \texttt{Stack-Fit} (S-Fit) technique from \cite{sfit} is used to remove PSF effects on the coadd level. For each individual exposure, for each CCD, the PSF is modeled by fitting to high signal to noise stars. The stars are decomposed into a linear combination of their principle components (eigen PSF). \cite{dls2d} show that 20 such components are sufficient to achieve an accurate model. A 3rd order polynomial is used to interpolate between modeled stars to create a spatially varying PSF on each chip for each individual visit. To construct a PSF on the coadd, the principal components for each individual exposure are stacked. Shape measurement is then carried out on coadded images by fitting stacked-PSF-convolved elliptical Gaussians to galaxies to estimate their semi-major and minor axis and orientation angle.  

We work in the weak lensing limit, where the ensemble average of galaxy elipticities, $\langle \mathbf{\epsilon} \rangle$, is an unbiased estimator of the reduced shear, $\mathbf{g}$. Deviations from the true reduced shear, $\mathbf{g}_{true}$ and the observed reduced shear, $\mathbf{g}_{obs}$ are given by a first order correction  $\mathbf{g}_{true} = (1 + m) \mathbf{g}_{obs} + C$. \cite{jt} use realistic image simulations of the DLS observing conditions to empirically determine the quantities $m_{\gamma} = (1 + m)$ and $C$ as a function of $R$ band magnitude and find

\begin{equation}
    m_{\gamma} = 6 \times 10^{-4}(m_{R} - 20)^{3.26} + 1.036
\end{equation}

and $C$ to be negligible. Similar to \cite{choi}, the sample of galaxies for which we use shapes is $22 \leq m_{R} \leq 24.5$, which corresponds to $1.04 \leq m_{\gamma} \leq 1.12$. 

\subsection{Photometric redshift estimation}
\label{sec:bpz}
The publicly available Spectral Energy Distribution (SED) template based code Bayesian Photometric Redshifts (\texttt{BPZ}) \citep{coe} was used for photometric redshift estimation. A detailed account is presented in \cite{sam}, but an abbreviated discussion follows. The six default SED templates packaged with BPZ were empirically adjusted based on overlapping spectroscopic redshifts obtained from the Smithsonian Hectospec Lensing Survey \citep[SHeLS,][]{shels}. A type and magnitude dependent prior was fit in the manner of \cite{benitez}, and relied on two sources of training data. The SHeLS data were used to fit the prior for galaxies $m_R$ < 21. For fainter galaxies up to $m_i$ < 24, spectroscopic redshift data from VIMOS-VLT Deep Survey are used for training \citep{2005A&A...439..845L}. $R$ band magnitudes were used for the magnitude dependence of the prior, and the template dependence was marginalized over. After the templates and priors have been adjusted, BPZ takes galaxy magnitudes and the DLS filter transmission curves as input to calculate a $\chi^2$ 

\begin{equation}
    \chi^2(T, z, a) = \sum_{i} \bigg[\frac{f_{i} - af_{Ti}(z)}{\sigma_{f_{i}}} \bigg]^2
	\label{eq:bpz}
\end{equation}

where $f_{i}$ is the observed flux in the $i$th filter with error $\sigma_{f_{i}}$, $f_{Ti}(z)$ is the predicted flux for a template (SED) $T$ at redshift $z$ after convolution with the system throughput, and $a$ is a template normalization factor. $\chi^2$ values are calculated on a grid for all template sets for redshifts in the range $0 \leq z \leq 3$, and span of normalization factors. The likelihood is then $\mathscr{L} \propto \mathrm{exp}(-\chi^2 /2)$. After weighting by the prior, the template set is marginalized over to produce a probability density function (PDF), denoted as $p(z)$.

Validating the photo-$z$ performance requires an independent sub sample for which spec-$z$s are known. The PRIsm MUlti-object Survey (PRIMUS) provides us with such a distinct validation sample, as it overlaps with F5 \citep{primus}. The PRIMUS sub-sample is photometrically complete to $R$ = 22.8, and contains a sample of randomly observed fainter objects that is 30\% complete with 22.8 > $R$ > 24.0. The DLS photo-$z$'s achieve a typical scatter of $\sigma_{photo-z} = 0.06\,*\,(1 + z_{spec-z})$ with an outlier rate of 4\% beyond $0.2\,*\,(1 + z)$ \citep{sam}. The magnitude range for which templates were empirically refined, priors were trained, and validation was performed places constraints on the magnitude range for which we have high confidence in photo-$z$ fidelity. This will be expanded on in the following section where galaxy selection and tomographic bin assignment are explicitly defined.

We will rely on the PRIMUS sub-sample when we attempt to empirically correct the mean and variance of the $n(z)$s in \ref{sec:pz-width}. Because PRIMUS was reserved for validation, we will use it to ensure our corrections are indeed making the estimated $n(z)$ more representative of the true $n(z)$.

\begin{table}
    \caption{Summary of cuts and descriptive statistics of galaxy populations in 
    tomographic bins used for this analysis. Count is total number of galaxies in bin,
    $\langle z \rangle$ is the mean redshift of the bin, determined by
    integrating the stacked $p(z)$ of the bin. L is for lens sample and S for source sample}
    \centering
    \begin{tabular}{lcccr} 
    \hline
    bin & z edges & $\langle z \rangle $  & $m_{r}$ & count\\
    \hline
    L0 & .37 - .48 & .422 & 20 - 22 & 24288\\
    L1 & .48 - .60 & .532 & 20 - 22 & 33918\\
    L2 & .60 - .80 & .679 & 20 - 22 & 29100\\
    S0 & .40 - .60 & .500 & 22 - 24.5 & 121491\\
    S1 & .60 - .80 & .693 & 22 - 24.5 & 88708\\
    S2 & .80 - 1.0 & .887 & 22 - 24.5 & 59845\\
    \hline
    \end{tabular}
    \label{Tab:bins}
\end{table}

\subsection{Tomographic selection}
\label{sec:tom}
At its heart, our work is concerned with the impact uncharacterized $n(z)$ errors have on cosmology estimation. The photo-$z$s used to estimate the $n(z)$ distributions for our samples must therefore be clearly defined. We discuss the selections used to define our sample here, paying mind to the measurements that will be made using these data.

In equations  \ref{eq:ggl} and \ref{eq:cluster} it is evident that the galaxy field and the convergence field are 'primary ingredients' in the two point probes. To directly measure these quantities, we must partition observed data into a \emph{lens} sample, where galaxy positions trace the galaxy field, and a \emph{source} sample, where galaxy shapes trace the convergence field. Below we discuss how these two samples are defined, and subsequently divided into tomographic bins. To begin, we discuss criteria that all galaxies used in this analysis must satisfy. Summary statistics that describe the tomographic bins are shown in Table \ref{Tab:bins}.

\cite{chrism} notes the 4 band filter set used for the DLS limits the reliable photo-$z$ range to $.4 \lesssim z \lesssim 1.0$. Consequently, our tomographic bins fall approximately in this range. To cull galaxies with unreliable photo-$z$s from the sample, we require galaxies are detected in all 4 photometric bands; the faintest galaxies are required to have $m_R \leq 24.5$, to allow sufficient signal to noise for robust photo-$z$s; the brightest galaxies allowed are $m_R \geq 20$ to exclude sources bright enough to saturate or have non-linear responses on CCDs. The bright cut also protects against a population of bright M-type stars, a contaminant which can masquerade as red galaxies in the 0 $ \leq z \leq $ 1 redshift range in DLS. The magnitude range for the experimental samples is representative of the magnitude range used to empirically adjust SED templates, train the prior, and validate photo-$z$ performance as discussed in \ref{sec:bpz}.

It is typical in tomographic analysis to use a point estimate-mode, mean, or median of the $p(z)$-to assign galaxies to tomographic bins \citep{cfht, abbot, troxel}. This can be problematic, as it allows for PDFs which are multi-modal, overly broad, or otherwise unreliable. To filter for more reliable photo-$z$ PDFs, we adopt the following technique. For a tomographic bin with edges $z_l$ and $z_h$, we integrate each galaxy's PDF in the interval $z_l - z_h$. If the integrated area is greater than 0.5  
its entire $p(z)$ is added to the tomographic bin. All such galaxy PDFs which satisfy this criteria are subsequently stacked to form an estimated $n(z)$ for that tomographic bin.

\emph{lens sample}: The magnitudes for all bins span $20 \leq m_{R} < 22$. To accommodate a source bin at higher redshift than the entire sample for the lens sample, we choose the total redshift range of $0.37 \leq z \leq 0.8$. This range is then subdivided into three tomographic bins. The width of the tomographic lens bins is largely set by the overall shape of the $n(z)$; the galaxy counts at high redshift drop off rapidly after $z \sim 0.6$. We seek some empirical guidance on how to partition the bins. We calculate the cumulative distribution function (CDF) of a BPZ prior, which is calculated using the magnitude distribution for galaxies with R band magnitudes from 20 to 22. We then partition the CDF in the range $0.37 \leq z \leq 0.8$ into three roughly equal bins. This ultimately yields lens bins with redshift edges [0.37 - 0.48), [0.48 - 0.6), [0.6 - 0.8].

Luminous Red Galaxies (LRGs) are often used to define a lens sample when calculating galaxy-galaxy lensing and galaxy clustering. This has several advantages: 1) the high mass of these objects creates higher signal to noise tangential shear measurements, given the same source plane. 2) red eliptical galaxies separate themselves from other galaxies in colour-colour space as their SEDs are redshifted. This makes them ideal targets for template fitting photo-$z$ codes, which can deliver accurate redshift estimates. 3) the galaxy bias is clearly defined for a particular galaxy sample \citep{redmagic}. 

 Exclusively using LRGs delivers results which pertain to very specific parts of the matter density field, however. Aside from the possibility of introducing a bias, it is scientifically compelling to use more egalitarian lensing selections to probe the dark matter field more generally. From equation \ref{eq:ggl} and \ref{eq:cluster} we can see the power spectrum for galaxy clustering and galaxy-galaxy lensing is defined by the normalized redshift distribution of galaxies and the galaxy bias. This provides freedom to define less restrictive lens cuts, provided the $n(z)$ can be estimated reliably. 
 
 The bright, low redshift, luminous galaxies in the DLS provide such a sample, which we use for our lens selection. These can provide a high shear signal for a more representative sample of galaxies in terms of e.g. their type and colour. Consequently, a broader, more generalized statement about matter fluctuations follows from measurements on these galaxies.

\begin{figure}
\includegraphics[width=\columnwidth]{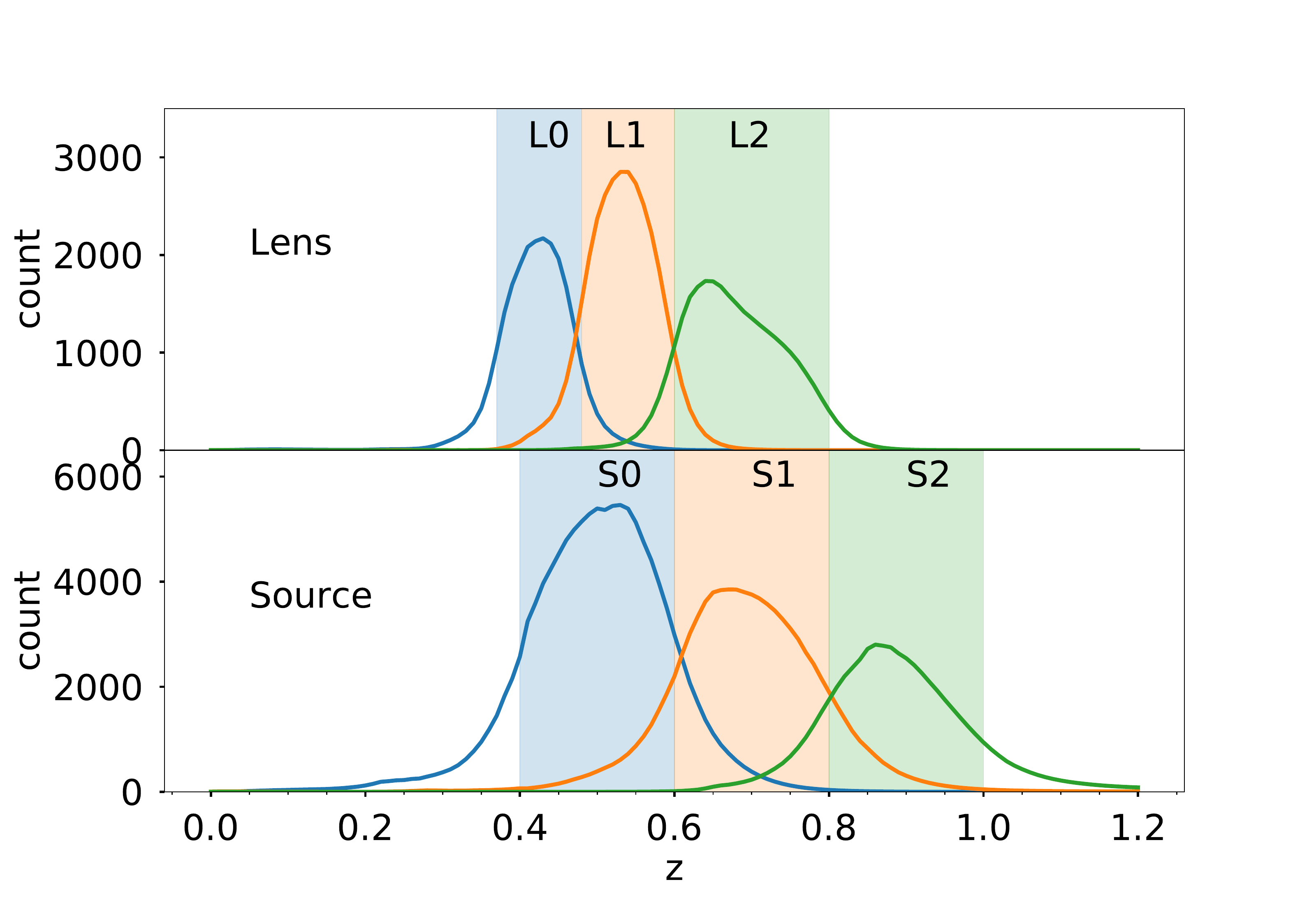}
    \caption{Redshift distributions for tomographic bins for the lens (top panel) and source (bottom panel) galaxies. Each solid curve shows the $n(z)$ of a particular tomographic bin. The solid vertical colours indicate the redshift ranges used to define the tomographic bins of their corresponding coloured $n(z)$ curves. $n(z)$ distributions within each bin are constructed by stacking the $p(z)$ distributions for galaxies which have 50\% of their integrated probability density within the redshift boundaries that define the bin.  Using this bin definition does result in distribution tails that extend beyond the bin boundaries. 
    }
    \label{fig:bins}
\end{figure}

\emph{source sample}: The magnitudes for all bins span $22 \leq m_{R} < 24.5$, which ensures that any one galaxy can only be placed in one lens bin or one source bin. To allow these galaxies to be lensed by the foreground galaxies in the lens sample, we choose three tomographic bins with redshift intervals [.4 - .6), [.6 - .8), [.8 - 1.0]. Accurate shapes are imperative for shear estimation. Thus, a prerequisite for source galaxies is that the elliptical Gaussian fitting procedure used to measure its shape successfully converged. We apply shape cuts to ensure the fitted ellipticities are reliable. The typical shape noise for the DLS is $\sigma_e \approx .35$, and so we require the shape error $\delta e < .25$ so that it is comparable to-and never dominates-the shape noise. In Figure 14  of \cite{dls2d}, the authors demonstrate strong ellipticity bias in small galaxies-regardless of signal to noise and shape error. This shape bias is due primarily to under sampling and pixelization effects, and can be avoided by requiring the semi-minor axis $b$ measured from the elliptical Gaussian fitting procedure is larger than 0.4, which we enforce on all source galaxies.

In Figure \ref{fig:bins} we show the resulting $n(z)$ distributions for both the lens samples (top panel) and source samples (bottom panel). $n(z)$ distributions for L0, L1, and L2 are shown as the blue, orange and green curves in the top panel respectively, and the extent of these bins are indicated by the shaded regions of the same colour. Similarly, the $n(z)$ distributions for S0, S1 and S2 are shown in the bottom panel as the blue, orange, and green curves respectively, and the shaded regions indicate the width of the bins or corresponding colour.

The various $n(z)$ shapes for our tomographic bins are complex and influenced by the overall $n(z)$ of the entire galaxy sample from the DLS. Selection criteria (integrated area inside of a bin, magnitude and shape cuts) additionally influence the shape of these curves. 

\section{Measured Two Point Statistics}
\label{sec:meas}
\subsection{Galaxy clustering}
If the likelihood of finding a galaxy inside a small patch on the sky subtended by solid angle $\Omega$, is $P_{\Omega}$, the likelihood of finding a neighboring galaxy an angular distance $\mathbf{\theta}$ away is given by $P_{\Omega}^2[1 + w(\theta)]$. $w(\theta)$ is the probability in excess of random correlations, and is expected to be non-zero when measured on galaxies which live in the same LSS. To measure $w(\theta)$ between two fields $i$ and $j$ we use the Landy-Szalay estimator \citep{ls}

\begin{equation}
    w(\theta) = \frac{\langle D_i D_j \rangle + \langle R_i R_j \rangle - \langle R_i D_j \rangle - \langle D_j R_i \rangle}{\langle R_i R_j \rangle}
\end{equation}

where $D_i D_j$ is the cross correlation between data in the $i$th and $j$th fields, $R_i R_j$ is the cross correlation of the $i$th and $j$th random fields, and $D_{i/j} R_{j/i}$ is the cross correlation between the $i/j$th Data field and $j/i$th random field. The case where $i = j$ yields auto-correlations. For every DLS field, for every galaxy field tomographic bin, we generate 7 times as many randoms as data points. Having randoms that accurately sample the survey's selection function is crucial; the random field's point density is compared directly to the data field's point density in order to establish and detect excess clustering of the data.

$w(\theta)$ is sensitive to masking effects, and having accurate masks and boundary regions which match the data and simulated random points is crucial. For our lens sample and generated randoms, we use the masks previously developed for the DLS\footnote{http://dls.physics.ucdavis.edu/imageaccess.html}.
As \cite{chrism} indicates, the DLS has low signal to noise regions between subfields and along the edges of the field footprints due to the dithering pattern used. The lens sample defined for this work, however, is comparatively bright, going as faint as $m_R = 22$. At this magnitude range, the survey is depth complete, even in the lower signal to noise regions. As \cite{yoon} point out, for such a conservative magnitude cut, we may neglect higher order corrections due to lower signal to noise, extinction and so forth. 

Because of the size of the DLS fields, our measurements of $w(\theta)$ rely on galaxies in five highly localized sky regions. $w(\theta)$ measurements, consequently, will be biased with respect to the true, global, value of $w(\theta)$. This deviation can be corrected by adding the integral constraint \citep{1980lssu.book.....P}.
The additional offset is typically calculated by minimizing the chi square of a power law model fit to the measured $w(\theta)$ signal \citep{yoon}, where the initial guess of model parameters are motivated by a cosmological model. \cite{matthews} noted that angular bins for $w(\theta)$ are known to be correlated, and consequently knowing the co-variance of $w(\theta)$ is necessary to obtain an accurate model fit. A complete discussion of the full co-variance matrix-including the covariance of $w(\theta)$ measurements-of both two-point statistics is presented later in section \ref{sec:cosmo}. However, we will note here for clarity that in essence the covariance is estimated by calculating the covariance of many measurements of  $w(\theta)$ on simulated DLS realizations.

The fitting function is of the form $w(\theta) = A(\theta)^{1-\gamma} - C$ where $C$ is the integral constraint, $A$ is a constant, and $\gamma$ parametrizes the power law slope. \cite{matthews} also indicate that fitting all three parameters at once is highly degenerate. To obtain reliable fits for all three parameters, we fix $\gamma$ and fit for $A$ and $C$ by reducing the $\chi$ square. We then repeat this procedure for a vector of $\gamma$ values, and select the combination of parameters for which the $\chi^2$ is smallest. 

\begin{figure}
\includegraphics[width=\columnwidth]{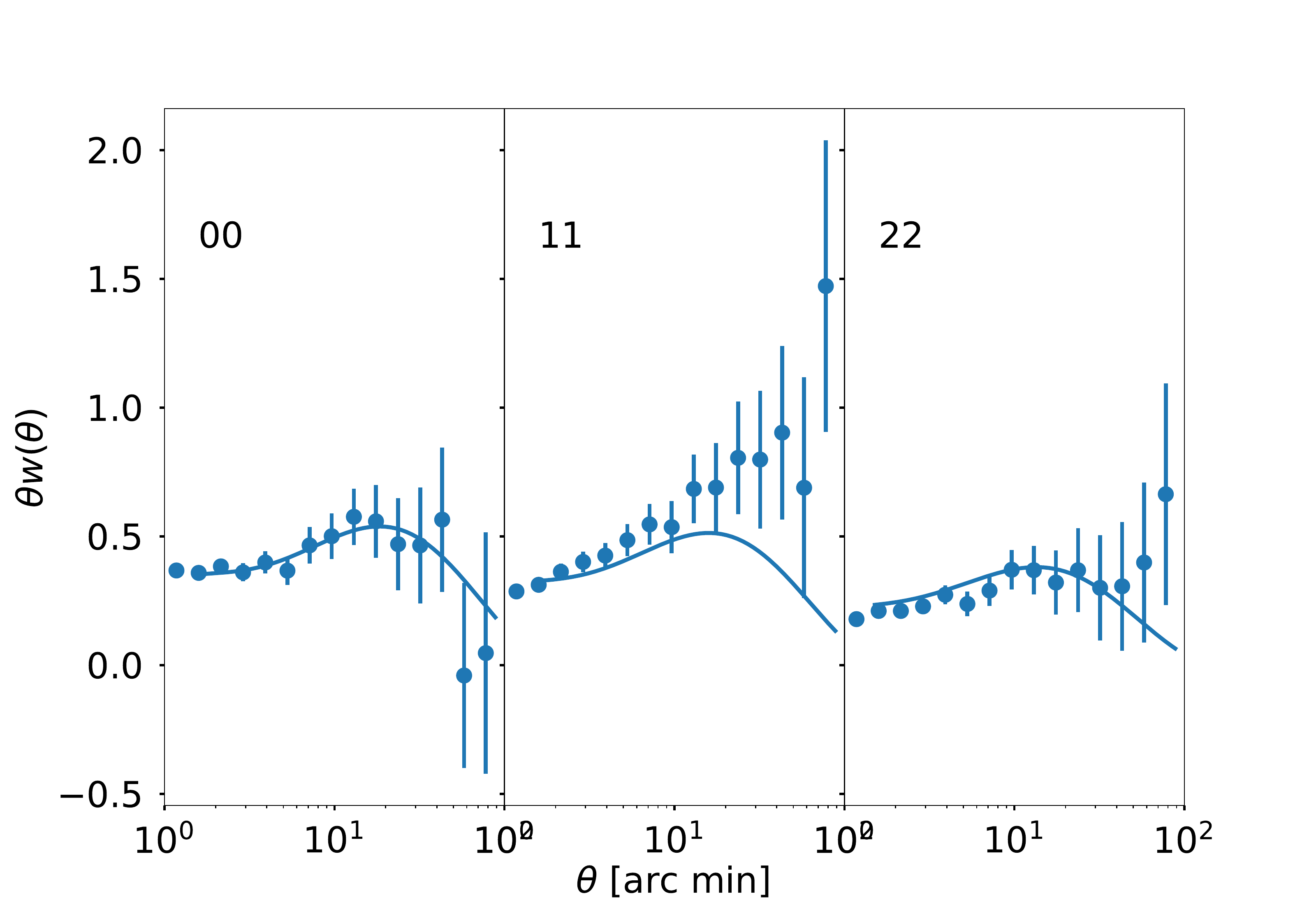}
    \caption{Measured Galaxy clustering auto and cross correlation function for three tomographic bins from the lens sample,
    [.37 - .48), [.48 - .6), [.6 - .8] in blue points. The solid lines show the theory prediction of the best fit cosmology from our cosmology inference. Number pairs in top right corner of the panels indicate which bins are being correlated, 
    e.g. 00 is the auto correlation of the (.37 - .48) bin. Because the points and errors are correlated, we caution the reader from doing a "chi-by-eye" fit.}
    \label{fig:w}
\end{figure}

The measured $w(\theta)$ signals for all auto-correlations of lens bins are shown in individual panels in Figure \ref{fig:w} as blue data points. 
The top left corner in each panel contains a pair of integers, which designates the lens bins being correlated (e.g. 00 auto correlates L0). The error bars on the data points come from the covariance matrix, which is estimated by realizing the DLS 549 times using simulations, the full details of which are discussed in section \ref{sec:cosmo}. Each panel also contains a solid line, which is the theoretical prediction from the best fit flat $\Lambda$CDM cosmology.

\subsection{Galaxy-galaxy lensing}
The gravitational field of massive foreground galaxies distorts the shapes of background galaxies, making them coherently tangentially aligned. Thus, measuring the shapes of background galaxies informs us of the matter distribution that host the foreground galaxies.

We first define the tangential and cross components of the ellipticities of background sources.

\begin{equation}
    e_t = -\Re[e\exp(-2i\phi)], \, \, e_{\times} = -\Im[e\exp(-2i\phi)], 
\end{equation}

where $e$ is the magnitude of the complex ellipticity, and $\phi$ is the azimuthal rotation angle which defines the source galaxy's position relative to the lens galaxy it is being correlated with. For an individual foreground lens, the galaxy-galaxy lensing signal is

\begin{equation}
    \gamma_{t / \times}(\theta) = \frac{\sum e_{i,t / \times} w_i}{\sum w_i}
\end{equation}

where $e_{i,t/\times}$ is the tangential/cross component of the complex ellipticity of the $i$th source galaxy from a tomographic bin, and $w_i$ is its weight. We define the per-galaxy weight as the inverse of the sum of the shape noise and shape error. True astrophysical signals are expected to have zero cross component signal absent the effects of source clustering and multiple lensing in cluster environments \citep{bradshaw}. The amplitude of the cross signal can, however, be used for a null test for systematic errors, and motivate scale cuts.

We then stack up the tangential and cross signal across galaxies in each lens-source tomographic bin combination. Finally, a field geometry correction is applied by subtracting the tangential signal around randomly placed points \citep{singh}. This ultimately makes our galaxy-galaxy lensing estimator 

\begin{equation}
   \langle \gamma_{t / \times} \rangle = \langle \gamma_{t/ 
   \times}(\theta)_g \rangle - \langle \gamma_{t / \times}(\theta)_r \rangle
\end{equation}

where $g$ $(r)$ is for the galaxy (random) field.

\begin{figure}
\includegraphics[width=\columnwidth]{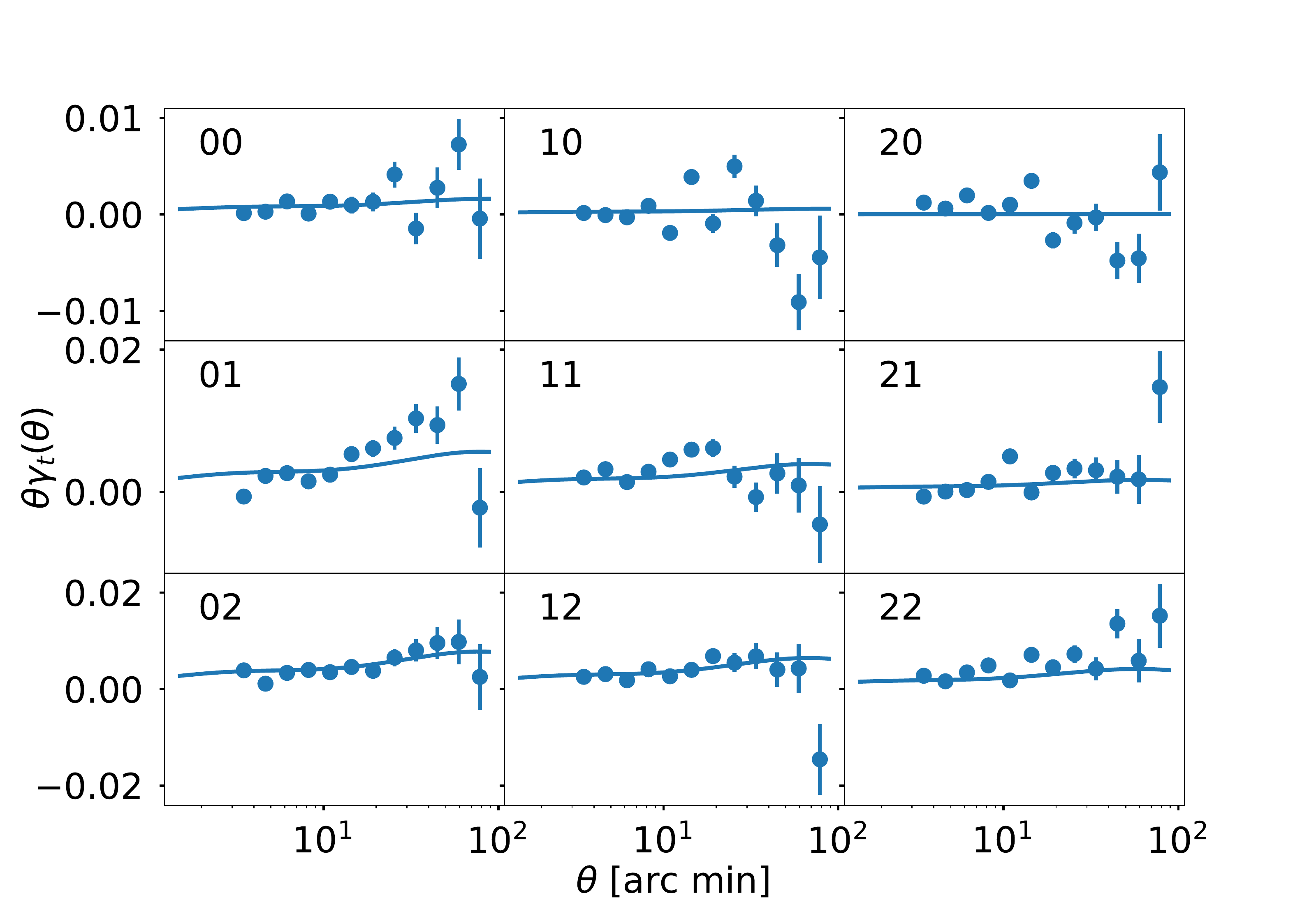}
    \caption{Measured galaxy-galaxy lensing signal (blue points) where positions from the lens sample
    are correlate with shapes from the source sample. The solid lines show the theory prediction using the best fit cosmology from our cosmology inference. Numbers in the upper right indicate which lens bin (left number) is correlated with which source bin (right number). Again, we caution the reader from doing a "chi-by-eye" fit since the data points and their errors are correlated.
    }
    \label{fig:gammat}
\end{figure}

The measured tangential shear signals are shown in Figure \ref{fig:gammat} as solid blue points. The top left corner of each panel contains two integers which indicate the lens and source bins are being correlated together (e.g. 00 shows where the galaxy positions of L0 cross correlate with galaxy shapes in S0, and 12 shows L1 positions correlated with shapes from S2. Note, when the first integer is larger than the second, the source population's mean redshift is lower than the lens population's. Depending on the degree of overlap of the lens and source bins, we anticipate small or no lensing signal in the ideal case.

\section{Corrections to $n(z)$ first and second moments}
\label{sec:correct_nz}
Validating the estimated redshift disrtibution derived from photometric redshifts for a galaxy sample requires knowledge of the true $n(z)$ of that sample. The SHELS and VVDS data provide a sample of spec-$z$s, however they are used to tune template sets and train the prior for our BPZ analysys. Consequently, they cannot provide unbiased metrics for photo-$z$ performance.  Instead, we use galaxies with secure spectroscopic redshifts from the PRIsm MUlti-object Survey \citep[PRIMUS][]{primus}, which were not used to train the photo-$z$s.  PRIMUS partially overlaps with F5 
but does not cover the DLS footprint completely. Because SHELS and VVDS overlaped with F2, PRIMUS provides a distinct sample in a region on the sky widely separated from training galaxies. PRIMUS is depth complete to $R = 22.8$ and additionally observed 30\% of galaxies with magnitudes $22.8 \geq R \geq 23.5$, which were chosen at random. Thus, the lens sample has complete coverage in terms of apparent magnitude, while the source sample is incomplete in terms of depth and magnitude range. We also note that properties of galaxies in F5 (even galaxies that are depth complete) are not guaranteed to be fully reflective of galaxies in F1-F4. Recall that the fields are 2 square degrees, and widely separated on the sky; sample variance may make one field unrepresentative of the others.

\subsection{Photo-$z$ mean}
To assess the impact of error in the mean redshift of the stacked photo-$z$s, we define the $z-$shift parameter, $\Delta z$, as the difference between the true $n(z)$ mean and mean of stacked $p(z)$. By translating the estimated $n(z)$ by $\Delta z$

\begin{equation}
    p^i(z) \rightarrow{} p^i(z + \Delta z_i)
    \label{eq:pzmean}
\end{equation}
\citep{yoon, hsc} the estimated mean is made to match the true $n(z)$ mean for the training data selected in the same manner. We will define $\Delta z_i$ for the $i$th tomographic bin for both lenses and sources. Later in Section \ref{sec:cosmo} we will treat all of these parameters as nuisance parameters and marginalize over them when performing our cosmology inference.

\begin{figure}
	\includegraphics[width=\columnwidth]{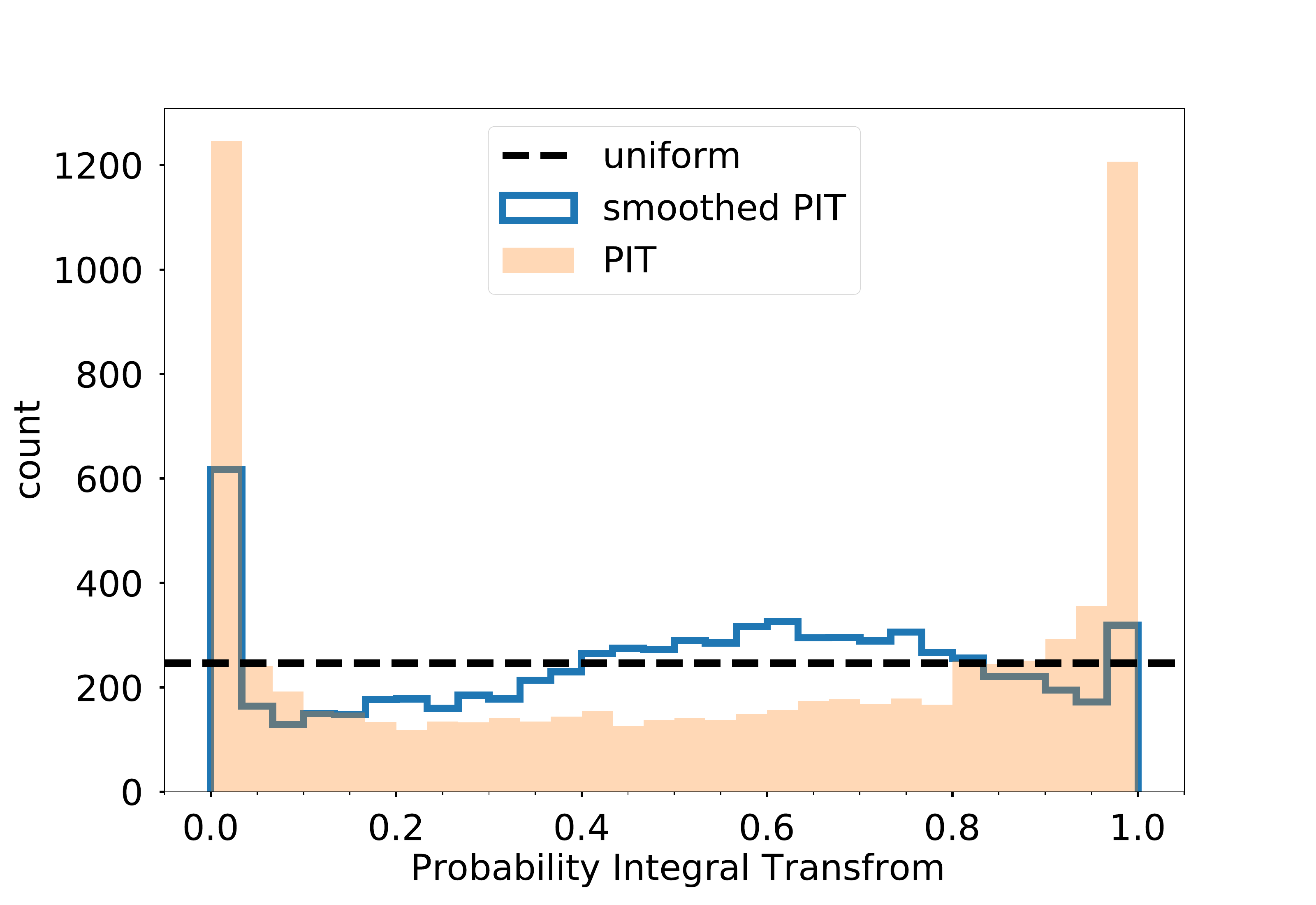}
    \caption{PIT histograms for galaxies in our sample for which we have an accompanying spec-$z$s from PRIMUS. For each galaxy in the DLS-PRIMUS sample, we calculate the cumulative distribution functions of their photo-$z$ PDF up to the true redshift. If the ensemble of PDFs were an accurate description of the true underlying $n(z)$, the histogram of all such values would be uniform (dashed black line). The orange distribution is the PIT histogram of DLS-PRIMUS galaxies before they are broadened with an optimal filter. The 'pile up' at the edges indicates the PDFs are overly narrow. The blue histogram shows the distribution after each individual PDF is convolved with a Gaussian. The width of the Gaussian kernel is calculated by splitting the galaxy sample into 4 roughly equal magnitude bins, and minimizing the Kullback-Leibler divergence for each magnitude bin. The optimal smoothing is able to broaden individual $p(z)$s 
    so they form a more representative description of the true $n(z)$ of galaxies considered. Regardless, a persistent bias is evident in the blue histogram as it trends upwards from left to right, indicating the right $p(z)$ tails truncate
    faster than the left tails, suggesting the $p(z)$s may not capture the $n(z)$.}
    \label{fig:pit}
\end{figure}

\subsection{Photo-$z$ width}
\label{sec:pz-width}
In equation~\ref{eq:cluster} the amplitude of the galaxy clustering power spectrum is not only affected by the linear galaxy biases, but also the tails of the tomographic bin redshift distributions. In effect, an error in the $n(z)$ width can imitate a cosmological signal that may erroneously be attributed to the galaxy bias or $n(z)$ mean. This is potentially problematic for cosmological inference: the joint combination of galaxy clustering and galaxy-galaxy lensing is meant to interlock such that the bias is constrained and removed. However, if degeneracies arising from $n(z)$ errors are unchecked, obtaining a high fidelity galaxy bias may not be achievable. It is therefore crucial to ensure that the characterization of overlap in the wings of the $n(z)$ is accurate. The ultimate impact on cosmology inference will be addressed in section \ref{sec:cosmo}.

We empirically adjust the widths of the $n(z)$ for each tomographic bins by adjusting the widths of the individual $p(z)$ for the galaxies that are included in said tomographic bin. To do so we require a metric to evaluate how faithfully the ensemble of $p(z)$s stacked together represents the underlying $n(z)$. The Probability Integral Transform (PIT) \citep{hscpz}
can be used to evaluate if the set of redshift PDFs for galaxies is a reasonable description of their true underlying $n(z)$. Formally, the PIT for a single galaxy, for which we have knowledge of the true redshift, is defined as

\begin{equation}
    \mathrm{PIT} = \int_{-\infty}^{z_t} p(z)\, \mathrm{d}z
	\label{eq:pit}
\end{equation}

where $z_t$ is the true redshift and $p(z)$ is a galaxy's PDF. 
If the $p(z)$ values accurately reflect the likelihood of observing a galaxy as a function of redshift, then the histogram of PIT values is expected to be consistent with a uniform distribution.  By extension, if the $p(z)$ are accurate, then we expect the stacked $n(z)$ estimate to also be accurate.
If the PDFs are overly broad, the PIT histogram is expected to have a peak near its center and a deficit of values at the extremes. On the other hand, if the PDFs are overly narrow or suffer from many catastrophic outliers, the PIT histogram will 'pile up' at the limits of the histogram range and have a deficit of values near 0.5. Calculating the PIT requires knowledge of the true redshift, so we are restricted to calculating it for the sub sample of DLS data that have spec-$z$s from PRIMUS.

In Figure \ref{fig:pit} we show the PIT histogram of the PRIMUS-DLS sample in solid orange before any adjustments are made to individual photo-$z$ PDFs. The build up at the limits of the PIT histogram is indicative of overly narrow PDFs. One possible explanation is that the photometric errors used in BPZ were underestimated; \texttt{Source Extractor} used for flux and flux error measurements in DLS processing, and was suspected of underestimating photometric errors. Additionally, \cite{davepz} demonstrate that absence of uncertainty in BPZs SED templates lead to overconfident $p(z)$s.  Under-predicted and unaccounted photometric errors subsequently yield overly confident individual PDFs. 

Following the method of \cite{hoyle} and \cite{davepz}, we posit that the widths of the stacked photo-$z$ PDFs can be made more reflective of their $n(z)$ by convolving their individual photo-$z$ PDFs by a Gaussian filter

\begin{equation}
    p(z)` \xrightarrow{} N(\sigma) \otimes p(z)
	\label{eq:convolve}
\end{equation}

where $N(\sigma)$ is a Gaussian with width $\sigma$ . To find an optimal width, we minimize the Kullback-Leibler divergence between the PIT and uniform case. 
For two discrete distributions, $P(x)$ and $Q(x)$, the Kullback-Leibler divergence is defined as 

\begin{equation}
    \sum_i P(x_i) \log[P(x_i)/Q(x_i)]
\end{equation}

and quantifies the difference between the distributions. We let $Q$ be a uniform distribution, and fit for a Gaussian width which minimizes the divergence when convolved with the photo-$z$ PDF before the PIT is calculated. A Kullback-Leibler divergence value of 0.0 in this case would indicate the PIT histogram is identical to a uniform distribution. We partition our spectra sample into 4 bins by $R$ band magnitude, where each bin has roughly the same number of spec-$z$'s. The bins are [20 - 21.65), [21.65 - 22.27), [22.27 - 22.7), and [22.7 - 23.5]. The optimal $N(\sigma)$ widths obtained for the individual training bins are 0.045, 0.0695, 0.0896, and 0.128 respectively.
Convolving the remaining $p(z)$ curves with their appropriate $N(\sigma)$ filter results in a Kullback-Leibler divergence for the whole sample of 0.053, a decisive improvement over the 0.276 before convolution.
The resulting PIT histogram from broadening PDFs is shown in Figure \ref{fig:pit} in the unfilled blue histogram. The post-colvolution PIT histogram is much closer to the uniform distribution (dashed black line) than the pre-convolved PIT histogram (solid orange), consistent with our Kullback-Leibler divergence calculations.

Additionally, we train the widths for each individual tomographic bin. The per bin training yielded a Kullback-Leibler divergence of 0.0627. Aside from being sub-optimal to the magnitude training, this raises concerns about over-training. The widths for bins in the F5 sample may be completely different than those in the other fields. A common technique to check for over training is to set aside part of the data for validation, and part for training. The dearth of galaxies with spec-$z$s in this sample makes this pursuit impractical. Splitting the sample would produce lower fidelity fits in addition to giving noisy validation results. Thus, we proceed using convolution widths from the magnitude training. 

Once we have determined the optimal filtering width as a function of $R$ band magnitude, we iterate over our PDFs, convolving them with their appropriate Gaussian filters according to the training model. The individual broadened PDFs are then assigned to tomographic bins such that their bin membership is identical to that before the PDFs were broadened. Finally, PDFs in the same bins are summed to create an estimated $n(z)$ for each tomographic bin.

\begin{figure}
	\includegraphics[width=\columnwidth]{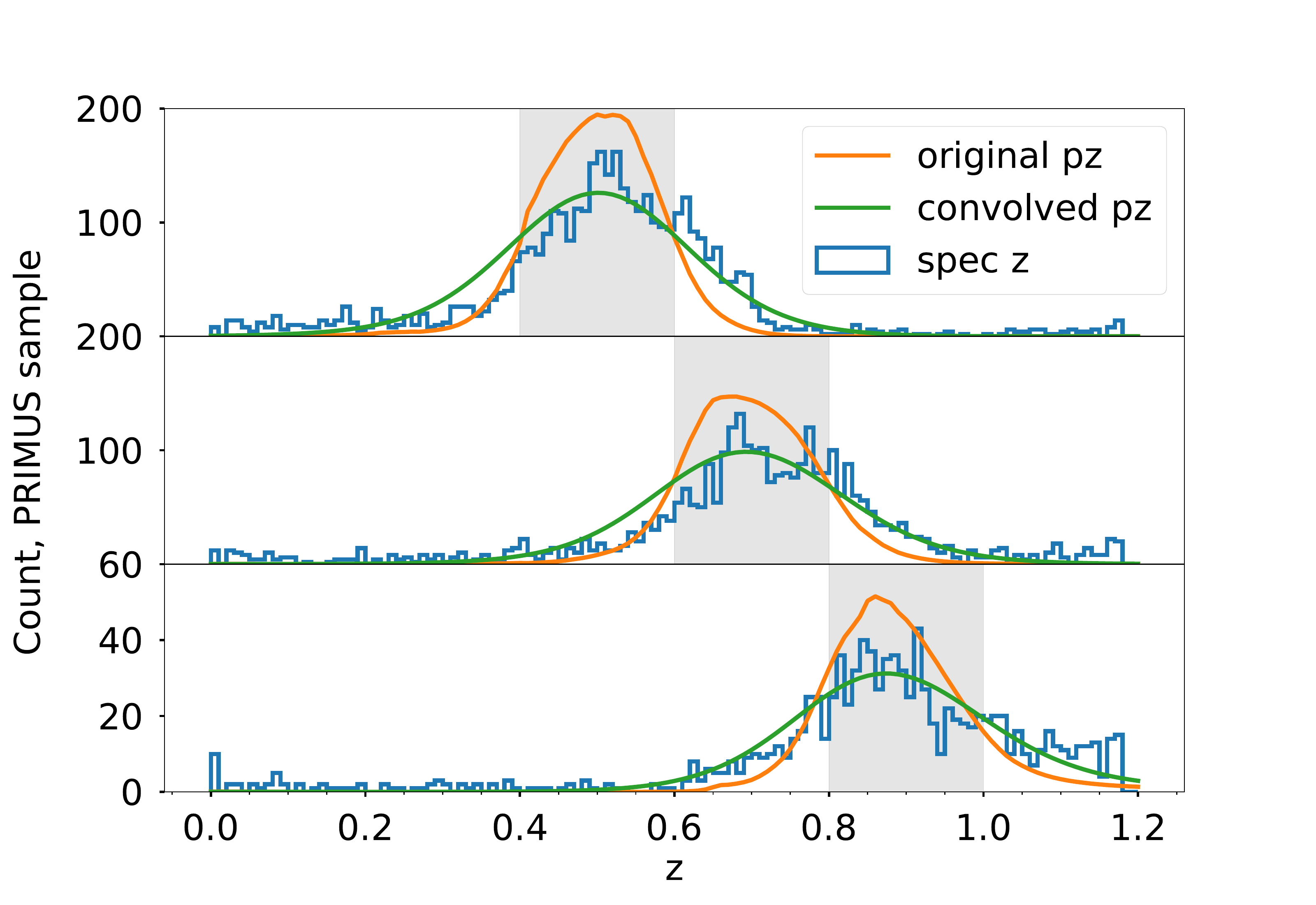}
    \caption{
    Comparison of the $n(z)$ distributions for a sub sample of DLS galaxies for which spec-$z$s are known.
    Galaxy selection and tomographic binning follows the procedure outlined in \ref{sec:tom}. Solid gray shaded
    regions indicate the redshift range that defines the tomographic bins. From top 
    to bottom, the panels are for Source bins 0, 1, and 2. The orange curves show the $n(z)$ from stacking
    $p(z)$s, as outlined in \ref{sec:tom}. The blue histograms show the corresponding $n(z)$ using spec-$z$s.
    Discrepancies between the $p(z)$ and spec-$z$ $n(z)$ mean and width are evident by eye.
    We attempt to mitigate the differences by convolving individual $p(z)$s with an optimal filter before stacking them, as discussed in \ref{sec:pz-width}, giving the solid green curves.
    The width correction techniques make the stacked $p(z)$ more true to
    the true underlying $n(z)$-as measured by the blue histogram-vary considerably from bin to bin. The
    impact these $n(z)$ errors have on cosmology inference outlined in \ref{sec:cosmo}}
    \label{fig:smooth_gal}
\end{figure}

In Figure \ref{fig:smooth_gal} we show the effect of stacking fiducial $p(z)$s (orange curve), stacking broadened $p(z)$s (green curve) and the true underlying $n(z)$ using spec-$z$s (blue curves) for three different tomographic bins. Differences in the means of the curves are evident by eye. Additionally, the overall shapes-and notably the widths-of the stacked fiducial $p(z)$s differ considerably from the true distributions. Although this PRIMUS sample is not necessarily representative of the entire DLS sample, the differences in the means and widths of the inferred $n(z)$ obtained by stacking fiducial $p(z)$s and the true $n(z)$ are substantial, and may potentially introduce systematic errors. The extent that the mean and width errors in tomographic $n(z)$ bins impact cosmology inference will be examined in the following section.
\section{Joint Analysis and Cosmology Inference}
\label{sec:cosmo}

\subsection{Co-variance estimation}
Cosmology inference requires a covariance matrix in addition to the data vector. Calculating the covariance for the cosmological probes is still an active area of research \citep[see][and references therein for a full summary]{krause}. 
There are several families of techniques that have been used in the literature to obtain a robust covariance matrix in weak lensing studies. These include internal estimators, analytically calculated covariances, and numerical simulations. Internal estimators like the Jackknife and Bootstrap technique re-sample the observed data set in order to calculate covariances. This allows for effects like depth, field geometry, field masking, shape noise, and other survey specific systematics to be considered while calculating the covariance. Jackknifing has been found to be accurate provided the scales of the jackknife regions are the size of or smaller than the scales being probed. 
There are competing needs when using the jackknife: on one hand we have the need for the number of jackknife regions to exceed the number of data points, and on the other we would like to make the regions smaller to allow for more regions \citep{singh}. Indeed, the small survey footprint of the DLS makes jackknifing intractable. The small area does not permit a sufficient number of jackknife regions while simultaneously keeping the regions large enough to be bigger than the angular scales being considered by our two point probes.

Analytic covariances often involve the assumption of a Gaussian density field. This assumption is valid in the linear regime on large scales. Analytic covariances are computationally less expensive to calculate and do not contain statistical noise the other two strategies are susceptible to \citep{krause}. However, because of the small, complex, footprint of the DLS and the small shape noise of the survey, a Gaussian covariance matrix may not be best suited to this survey \citep{chihway}. 

\begin{figure}
    \centering
    \includegraphics[width=\columnwidth]{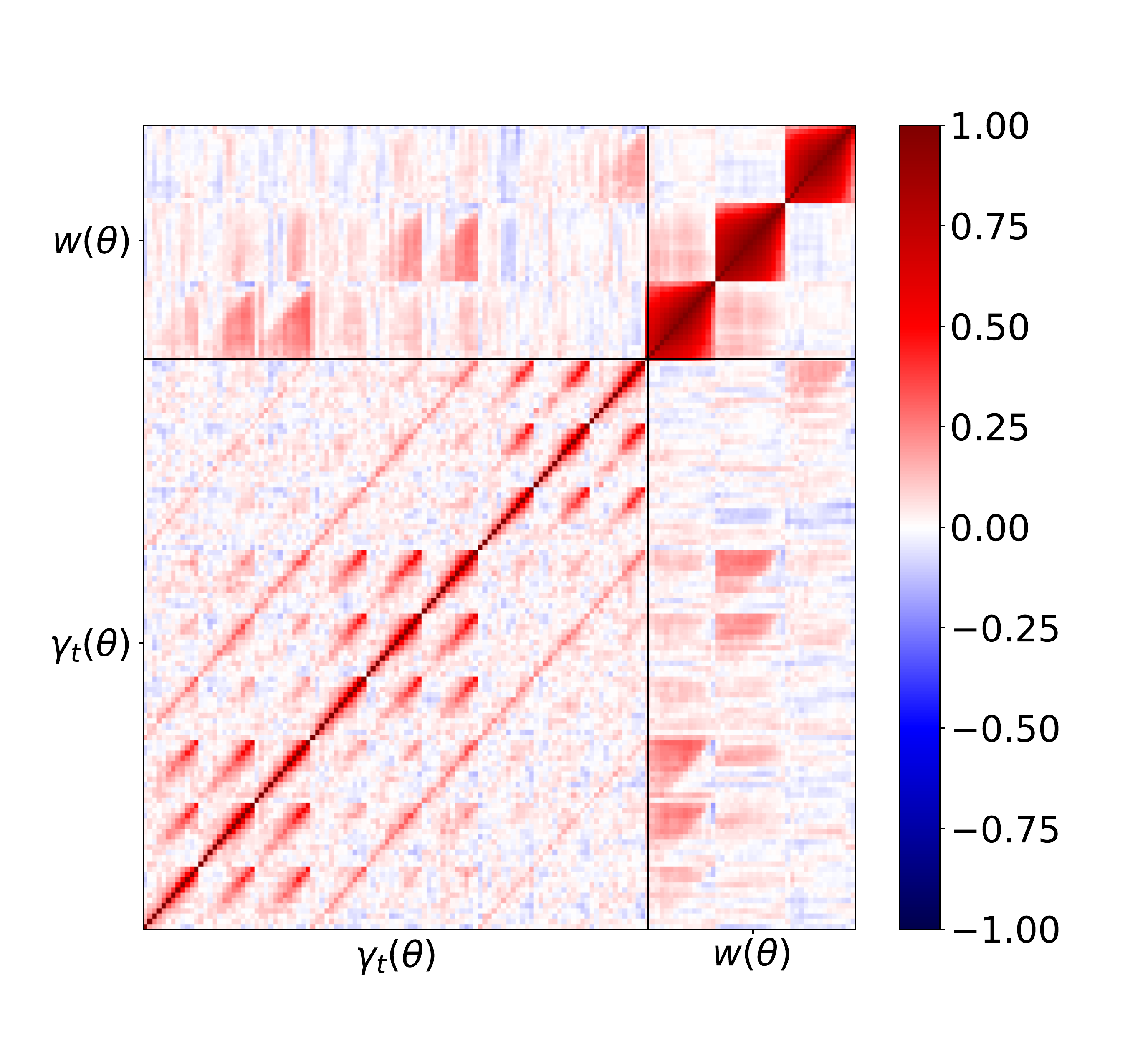}
    \caption{Estimated covariance matrix for the DLS data vector, created by simulating the DLS 549 times and measuring galaxy clustering and galaxy-galaxy lensing on the simulations}
    \label{fig:covariance}
\end{figure}

Simulations can allow for some survey specific systematics-like field geometry, an approximate description of shape noise, exclusion regions due to bright stars etc to be considered. If many mock surveys are created, the covariance matrix can be obtained by calculating the covariance of the data vectors from each mock. A caveat to bear in mind is that super-survey modes will not be captured by these simulations. We use the publicly available package Full Lognormal Astro-fields Simulation Kit (FLASK) to simulate 549 mock realizations of the DLS \citep{xavier}. By providing input power spectra for all possible combinations of convergence and matter density fields, cosmological parameters, and angular and radial selection functions, FLASK can simulate the DLS at a catalogue level. The catalogues include galaxy angular position on the sky, redshift, and galaxy shape after gravitational lensing. This enables us to measure galaxy clustering, and galaxy-galaxy lensing, using the same pipeline used for the real DLS catalogues. The covariance of these data vectors is calculated and used as the covariance matrix for our cosmological probes. In Figure \ref{fig:covariance} we show the resulting covariance matrix we derive.

\begin{table}
    \caption{Summary of nuisance, astrophysical, and cosmological parameters we marginalize over in our likelihood analysis. The photo-$z$ mean parameters are determined by comparing photo-$z$'s from subset of our data for which we we have spec-$z$s. Photo-$z$ biases have Gaussian priors, and remaining parameters have uniform priors. For Gaussian priors, the third column represents the mean and standard deviation of the prior. For uniform priors the column represents the edges of the flat tophat prior.}
    \centering
    \begin{tabular}{lr} 
    \hline
    parameter  & prior parameters\\
    \hline
    \multicolumn{2}{|c|}{\textbf{Nuisance Parameters}}\\
    L0 photo-$z$ bias  & 0 0.2 \\
    L1 photo-$z$ bias  & 0 0.2 \\
    L2 photo-$z$ bias  & 0 0.2 \\
    S0 photo-$z$ bias  & 0 0.2 \\
    S1 photo-$z$ bias  & 0 0.2 \\
    s2 photo-$z$ bias  & 0 0.2 \\
    shear calibration & -.03 .03 \\
    \hline
    \multicolumn{2}{|c|}{\textbf{Astrophysical Parameters}}\\
    I.A. Amplitude  & -5 5 \\
    Galaxy bias     & 0.8 3.0 \\
    \hline
    \multicolumn{2}{|c|}{\textbf{Cosmological Parameters}}\\
    $\Omega_m$ matter density &  .05 .9 \\
    $n_s$ spectral index    &  0.8 1.2 \\
    $\Omega_b$ baryon density & .03 .06 \\
    $\sigma_8$ power spectrum normalization&  .20 1.4 \\
    $h$  Hubble parameter      &  .50 .85 \\
    
    \hline
    \end{tabular}
    \label{Tab:bins}
\end{table}

\subsection{Likelihood analysis}
Ultimately, we wish to obtain constraints on cosmological parameters that define a cosmological model. We will proceed by using a Bayesian framework, which will allow us to produce a posterior distribution for cosmological parameters of interest. Formally, the posterior is defined as 

\begin{equation}
    P(\mathbf{Y(\theta})|\mathbf{X}) = \frac{P(\mathbf{X}| \mathbf{Y(\theta)})P(\mathbf{Y(\theta}))} {P(\mathbf{X})}
\end{equation}

where $\mathbf{Y}$ is the a model with parameters $\theta$, $ \mathbf{X}$ is the observed data, P($\mathbf{X}$|$\mathbf{Y}$($\theta$)) is the likelihood, $P(\theta)$ is the prior, and P($\mathbf{Y}$($\theta$)|$\mathbf{X}$) is the posterior

In principle one can use this framework to calculate posteriors for parameters in different cosmological models. Indeed, \cite{andy} argue that a joint combination of several cosmological probes is a promising avenue to measuring dynamic dark energy and modified gravity. Here we will restrict our attention to the case of flat $\Lambda$CDM.

Given a model cosmology with parameters $\mathbf{\theta}$ the likelihood function is given as

\begin{equation}
    L = \frac{1}{(2\pi)^{1/m}|C|^{1/2}} \mathrm{exp}[ -.5 (\mathbf{X} - 
    \mathbf{Y(\theta}))^\mathrm{T} \mathbf{C}^{-1} (\mathbf{X} - \mathbf{Y(\theta})) ]
\end{equation}

where $\mathbf{X}$ is the data vector, $\mathbf{Y(\theta)}$ is the predicted data value assuming a $\Lambda$CDM cosmology, $m$ is the dimension of the data vector, and $\mathbf{C}$ is the covariance matrix. In practice, it is more practical to use the log likelihood to avoid computational problems, and this is true in our work as well. As a consequence, $P(\mathbf{X})$ will become an additive constant we may neglect.

We will consider two cases, the uncorrected case where the model has the fiducial $n(z)$ bins, and the corrected case, where the model has our broadened $n(z)$ bins. We will then compare the constraints obtained in both cases and examine them for significant differences to assess if $n(z)$ widths errors can introduce systematic errors. 

We use the publicly available code \texttt{COSMOSIS} \citep{2015A&C....12...45Z}
to carry out a Monte Carlo Markov Chain to sample the posterior. At a high level, the predicted data vectors $\mathbf{Y}$ are generated by first assuming values of $\mathbf{\theta}$, which are sampled from their priors. The $\mathbf{\theta}$ are used to calculate the full three dimensional non-linear matter power spectrum using \texttt{CLASS} \citep{CLASS}and \texttt{HALOFIT} \citep{HALOFIT}. After, the limber approximation is applied to the 3 dimensional matter power spectrum to make it two dimensional.
The power spectrum of the different two point statistics (see eq \ref{eq:ggl}, and \ref{eq:cluster}) are derived from $P\Big(\frac{l + 1/2}{f(\chi)}\bigg)$ and depend on the normalized $n(z)$ of the tomographic bins being considered. A Hankel Transform is applied to the power spectra to generate prediction values in real space, which are arranged to form the predicted data vector $\mathbf{Y(\theta)}$.

\subsection{Priors}
We define several nuisance parameters which we marginalize over during our likelihood analysis. We use Gaussian priors for all $n(z)$ priors, defined by their mean and standard deviation. The mean and standard deviation are set to 0 and 0.02 respectively, for these priors. The standard deviation are determined by comparing a sample of photometric redshifts from galaxies which also have spectroscopic redshifts from PRIMUS, where the typical scatter is found to be approximately 0.02 \citep{sam}. 

The multiplicative galaxy bias, $b_i$ for each lens bin is of particular importance to our study. In addition to its high degeneracy with $\Omega_m$ and $\sigma_8$ \citep{2018PhRvD..98d2005P}, we are examining any degeneracy it may have with $n(z)$ shape errors. We use uniform priors from 0.8 to 3.0 for each galaxy bias. The large range in prior values avoids any prior bias in our likelihood analysis.

We use a flat prior in the range of -0.03 to 0.03 to marginalize over shear calibration errors with a uniform prior, following the method of \cite{yoon} and \cite{dls3d}. The multiplicative shear bias for the $i$th source bin, $m_i$, is defined as $\gamma_{t}^{'} = (1 +m_i)\gamma_{t} $

We also define several astrophysical parameters to marginalize over. Intrinsic alignments of galaxies can mimic a coherent alignment of galaxy shapes, contaminating the shear signal, and are a key systematic is weak lensing studies \citep{rachel}. We adopt an intrinsic alignment model which has amplitude dependence. We marginalize over $A$ with flat priors in the range -5 to 5.

Finally, we marginalize over cosmological parameters, all with uniform priors which encompass values obtained from recent studies. The ranges for these flat priors are 0.1 $\leq \Omega_m \leq$ 1.0; $0.5 \leq h \leq 0.85$; $0.2 \leq \sigma_8 \leq 1.4$; $0.8 \leq n_s \leq 1.2$; $0.02 \leq \Omega_b \leq 0.06$ 

\section{Results}
\label{sec:results}
\subsection{Effect of broadening $n(z)$}
\begin{figure}
    \centering
    \includegraphics[width=\columnwidth]{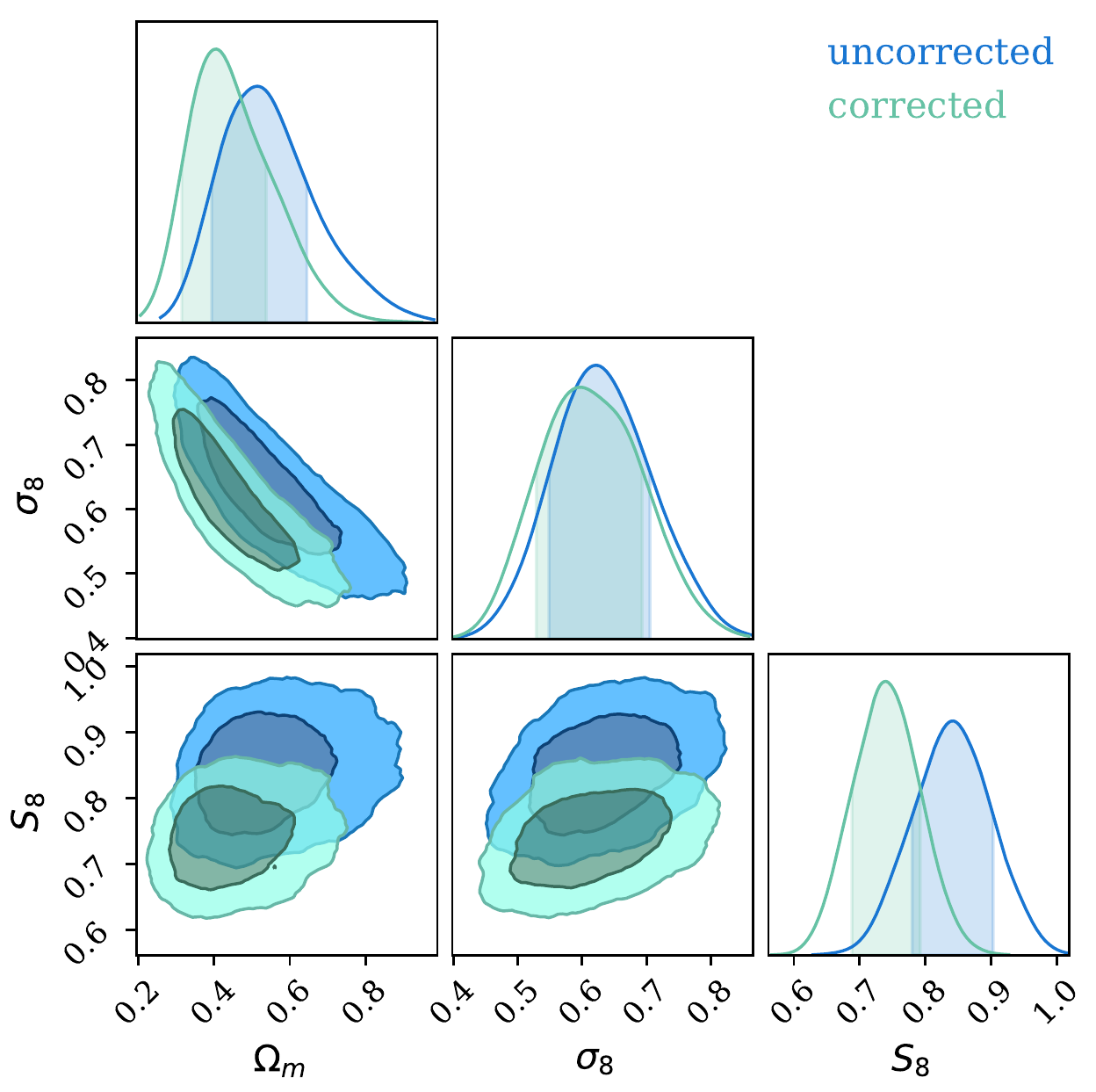}
    \caption{We present two cosmological constraints from combined galaxy clustering and galaxy-galaxy lensing probes. The uncorrected case, in blue, shows constraints where we have not attempted to correct the widths of the $n(z)$ distributions in our samples. The corrected case, in cyan, shows constraints after we have corrected the widths of the $n(z)$ distributions in our tomographic bins. The two cases are consistent at the 1$\sigma$ level, although the result of correcting the $n(z)$ is for the confidence interval to 'slide down' in the $S_8 - \Omega_m$ plane, ultimately shifting the constraint on the growth of structure parameter $S_8$ to slightly lower values.}
    \label{fig:broadened_banana}
\end{figure}

In Figure \ref{fig:broadened_banana} we show the parameter constraints on the cosmological parameters $\Omega_m$, $\sigma_8$ and the derived parameter $S_8 \equiv \sigma_8 (\Omega_m/.3)^{.5}$ for a flat $\Lambda$CDM cosmology using our DLS data. The degeneracy between $\Omega_m$ and $\sigma_8$ motivates the definition of $S_8$, which splits this degeneracy, as can be seen in the Figure. $S_8$ is also useful for comparing the results from different studies.

The results for the uncorrected set up, where we use the original $n(z)$ distribution before broadening with a trained filter, are shown in the blue contours where the darker region shows the 1-$\sigma$ interval and the lighter shaded region shows the 2-$\sigma$ region. The results for the corrected case, where we have broadened the $n(z)$ in the method discussed in \ref{sec:pz-width}, are shown in cyan. For the corrected setup we arrive at the following constraints: $S_8 = 0.739^{+0.054}_{-0.050}$, $\Omega_m = 0.412^{+0.113}_{-0.094}$ and $\sigma_8 = 0.601^{+0.088}_{-0.067}$

\noindent The corrected and uncorrected results are consistent with one another at the 1-$\sigma$ level. 
It warrants noting, however, that the value of $S_8$ is lower for the corrected case. Particularly, the uncorrected case is in good agreement with the previous DLS results for cosmic shear $S_8 = 0.818^{+0.034}_{-0.026}$ \citep{dls2d} and the DLS results for power spectrum space galaxy clustering + galaxy-galaxy lensing $S_8 = 0.810^{+.039}_{-.031}$ \citep{yoon}. The corrected result is still consistent with the previous DLS results at the 1-$\sigma$ level. 

\subsection{Effect of angular scale cuts}
In our analysis we implicitly made a choice on our angular scale cuts. The minimum scale is determined by model uncertainties such as baryonic physics and the accuracy of the non-linear power spectrum.
The maximum scale is set by the footprint of the five fields in the DLS. In \cite{dls2d} and \cite{chihway} the effects of how angular scale cuts on the DLS cosmic shear data impacted the cosmological constraints are presented. For example, when using conservative angular scale cuts where the minimum angular scale corresponds to 1.3 comoving Mpc in each redshift bin, the DLS constraint moves to a large values of $\Omega_m$ and $S_8$ to greater than .8 and .9 respectively. 

With an eye towards this, we investigate the effect of angular scale cuts on our work presented here. We compare three cases where we have unified angular scales to all probes: the corrected case where the minimum angular scale is 2 arcminutes, an intermediate case where we take the corrected data vector and impose a minimum angular scale cut is 6 arcminutes, and a conservative case where we take the corrected data vector and impose a minimum cut of 20 arcminutes. All maximum scale cuts are set to 90 arcminutes.

\begin{figure}
    \centering
    \includegraphics[width=\columnwidth]{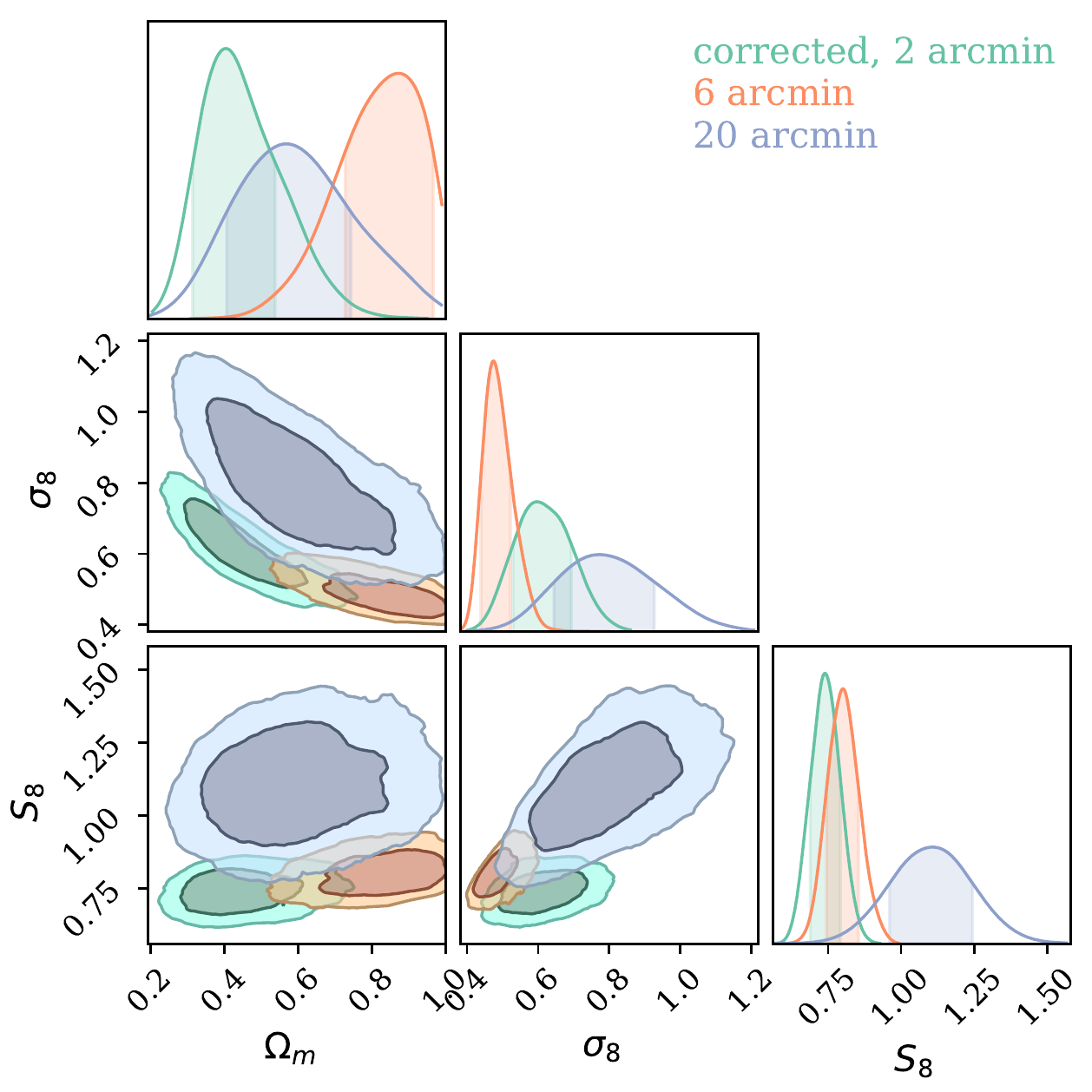}
    \caption{We present 3 cosmological constraints, each using different angular scale cuts in our galaxy clustering and galaxy-galaxy lensing data vector. The angular scale cuts are uniform across probes. Specifically, the corrected case (2-90 arcminutes), intermediate case (6-90 arcminutes) and conservative case (20-90 arcminutes) are shown in cyan, orange, and purple/blue respectively. The constraints are highly sensitive to angular scale cuts. Namely, different scale cuts on the same data vector produce different cosmological constraints. This likely indicates unresolved systematic effects in the data remain.}
    \label{fig:angle_banana}
\end{figure}

In Figure \ref{fig:angle_banana} we present the constraints of the different scale cut cases. The corrected case (2 - 90 arcminutes), intermediate case (6 - 90 arcminutes), and conservative case (20 - 90 arcminutes) are shown in cyan, orange, and purple/blue respectively. Several comments are in order. The conservative case noticeably loses constraining power relative to the other two cases. This is to be expected given the dramatic reduction in data that is being used in cosmology inference. Enlarged contours in the $\Omega_m - S_8$ planes are also presented in \cite{chihway} when conservative angular scale cuts are used on other surveys. However, in addition to broadening, the conservative case moves the contour up to higher values of $S_8$, exceeding 1. The intermediate case, on the other hand, seems to follow the direction of the 'banana' in the $\Omega_m - S_8$ plane, sliding down to high values of $\Omega_m$ and small values of $\sigma_8$. 

While the corrected case and intermediate case ultimately constrain $S_8$ values that are consistent with one another, we point out that the constrained values of $\Omega_m$ and $\sigma_8$ differ at the 1$\sigma$ level. Further more, the intermediate case gives values that are drastically different than those presented in the recent literature. 

As noted above, the conservative case presents as an outlier in the parameter $S_8$, constraining values between approximately 1 and 1.25 at the $1\sigma$ level. This value is in significant tensions with the other cases, and other values presented in the modern literature.

Ultimately, the key take away from Figure \ref{fig:angle_banana} is that different angular scale cuts lead to different cosmological constraints. This is a troubling observation, as it indicates some as of yet uncharacterized systematics still remain in the data. For a truly robust data set, the anticipated outcome of progressively moving the minimum angular scale to higher angles would be looser-but \emph{consistent} constraints.  That is, in the absence of uncharacterized systematics we expect larger error contours, but not the shifts in parameter values that we observe with different angular scale cuts. This implies that there are un-modeled scale dependent systematics in the data.

\section{Comparison of DLS $S_8$ with Other Studies}
\label{sec:compare_cosmo}
\begin{figure}
    \centering
    \includegraphics[width=\columnwidth]{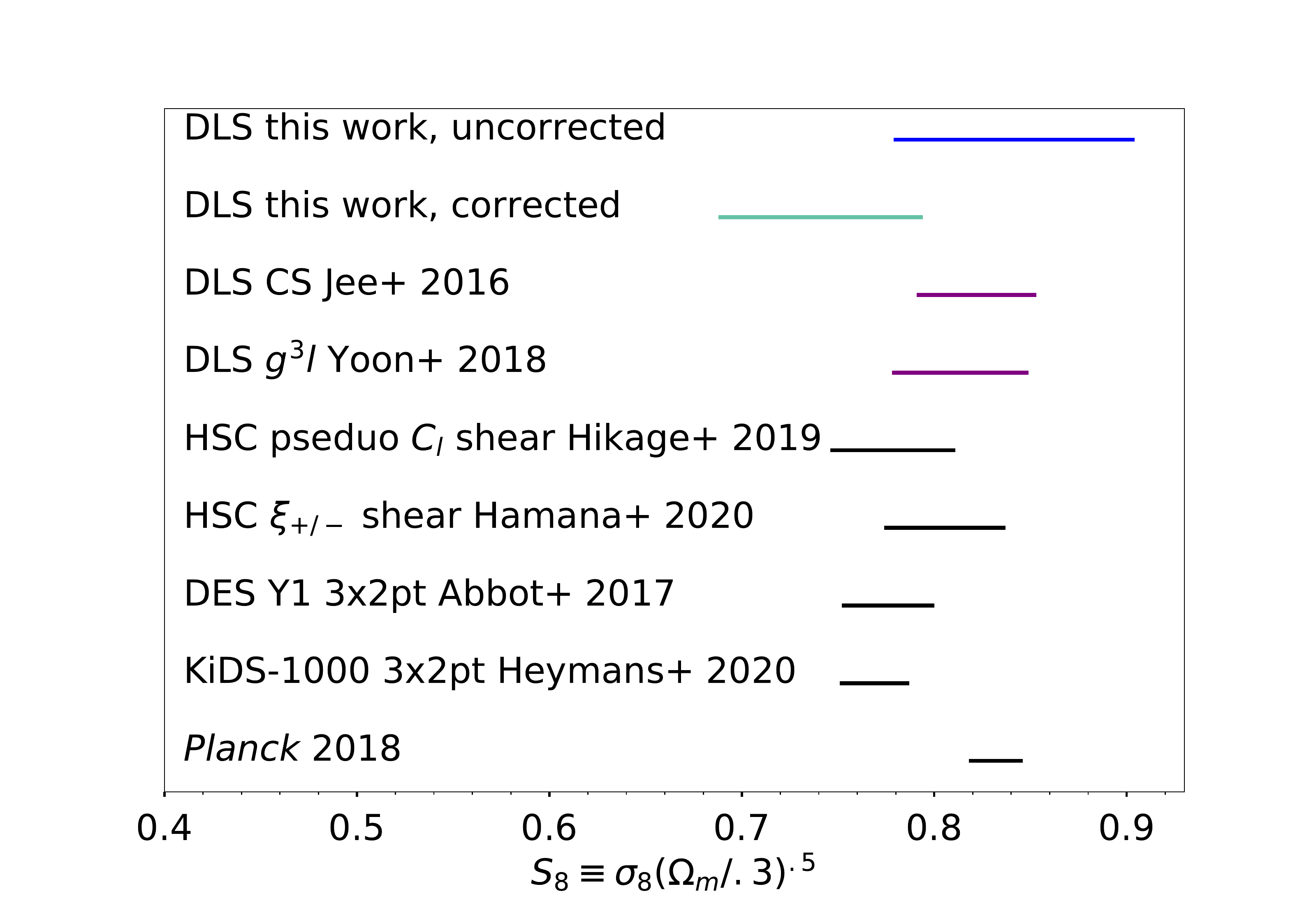}
    \caption{We show marginalized constraints on the growth of structure parameter $S_8 \equiv \sigma_8(\Omega_m/.3)^{.5}$ for our study, previous DLS cosmology studies, and recent constraints in the literature. While not as constraining as other studies, we nonetheless find agreement within $1\sigma$ of other weak lensing results and \emph{Planck}.}
    \label{fig:compare_cosmo}
\end{figure}

We compare our results to those of DES Year 1 \citep{abbot}, KiDS 1000 \citep{2020arXiv200715632H}, HSC Year 1 \citep{hsc, 2020PASJ...72...16H}, \emph{Planck} \citep{2020A&A...641A...6P}, as well as previous studies from the DLS. In Figure \ref{fig:compare_cosmo}, we show the resulting $S_8$ constraints from these studies. The first thing that draws the eye is the relative lack of constraining power in this work relative to the other studies presented. This is somewhat expected, as our especially conservative treatment of photo-$z$s lead to us discarding many galaxies in our sample. Our study relies on approximately a third of one million galaxies. Other DLS studies use approximately a million galaxies, and HSC, DES, and KiDS use several millions of galaxies, by comparison.

Overall we find good agreement between our study and other weak lensing studies. The lack of constraining power in our result, however, means that we cannot weigh in on the so called "$S_8$ tension", the tension in results for $S_8$ between some low redshift weak lensing surveys and \emph{Planck}. For example, in Figure \ref{fig:compare_cosmo}, the KiDS-1000 3x2 point result is in significant tension with the \emph{Planck} result, while the previous DLS results are in fair agreement with \emph{Planck}. The result presented here, on the other hand, has large enough uncertainty that it can straddle the \emph{Planck} result and other weak lensing results simultaneously. The shift between the DLS corrected and uncorrected $n(z)$ uses the same underlying model, whereas some of the other surveys shown marginalize over different parameters (such as feedback) or have different sample selections (angular cuts). It is also worth emphasizing that we use a different sample selection in this work compared to other DLS studies: the previous cosmic shear result utilizes a low $z < .4$ and high $z > 1$ redshift bin, and uses galaxies as faint as 27 in the $R$ band. The previous result from \cite{yoon} uses magnitude cuts similar to ours, but uses a low redshift bin that is recalibrated using overlapping spectroscopic redshifts from Sloan, and does not use the $n(z)$ width correction we discuss here. None the less, all DLS results are agree within $1\sigma$.

\section{Conclusions}
\label{sec:conclusion}
We utilize a Rubin Observatory LSST precursor, the Deep Lens Survey, to investigate how methods to mitigate errors on the $n(z)$ shape can impact cosmology constraints when using the combination of galaxy clustering and galaxy-galaxy lensing probes. To correct for the width of the $n(z)$, we use a validation set of data for which we have spectroscopic data, and empirically broaden individual photo-$z$ $p(z)$ with a best-fit Gaussian filter. The width of the Gaussian filter is selected by minimizing the Kullback-Leibler divergence between the probability integral transform of the photo-$z$ PDFs and the uniform distribution. We find that correcting the DLS $n(z)$ in this manner results in a shift in the constraint on the growth of structure parameter $S_8$ to lower values, though not at a statistically significant level. The dependence of cosmology parameters on correct knowledge of the tomographic bin widths shown in this analysis confirms the stringent requirements on these parameters shown in forecasts for Stage IV surveys \citep{ma}. While our result is not as constraining as other weak lensing measurements of $S_8$, we find good agreement with them and \emph{Planck} for our fiducial set of spatial scales. 

While our data set presents us with realistic photometric errors and photo-$z$ errors, they are also subject to many systematic effects, some of which may contribute to the $n(z)$ galaxy bias degeneracy. 
Truly understanding the impact that these degeneracies impose on cosmology inference requires they be examined in isolation of all other sources of error. This is perhaps intractable in experimental data and requires a realistic simulation to fully evaluate. This will be the focus of our next paper, where we will repeat the analysis presented in this paper on the Dark Energy Science Collaboration Data Challenge 2 truth galactic catalogue. Simulated data will also enable us to examine techniques to mitigate the effects of photo-$z$ width and skew in isolation of all other effects. Additionally, we may include a third constraint from adding cosmic shear together with our galaxy clustering and galaxy-galaxy lensing constraints. We do not expect the $n(z)$ width corrections to play a large role in the cosmic shear data, because of the broadness of the lensing kernel. Nonetheless, it is possible the inclusion of a third joint constraint may mitigate errors which arise in the galaxy clustering + galaxy-galaxy lensing constraint alone.

\section*{Acknowledgements}
This research was supported by DOE grant DE-SC0009999 and NSF/AURA grant N56981CC.
This work was performed in part at Aspen Center for Physics, which is supported by National Science Foundation grant PHY-1607611. Part of this work was performed under the auspices of the U.S. Department of Energy by Lawrence Livermore National Laboratory under Contract DE-AC52-07NA27344. IH would like to thank Perry Gee for help navigating DLS data and programming advice, and Henrique Xavier, Nickolas Kokron, Lucas Secco, Judit Prat and Mijin Yoon for their insights on using FLASK, and James Jee for his advice on using DLS data.

\section*{Data Availability}
The data underlying this paper was provided by the Deep Lens Survey collaboration. Data will be shared on reasonable request to the corresponding authors upon permission of the collaboration. The analysis and plotting code used to make this paper are publicly available on GitHub at the following URL: https://github.com/ih64/DLSJointProbes










\bsp	
\label{lastpage}
\end{document}